\documentclass[sn-mathphys,Numbered, icol]{sn-jnl}

\usepackage{graphicx}%
\usepackage{subcaption}
\usepackage{mwe}
\usepackage{multirow}%
\usepackage{amsmath,amssymb,amsfonts}%
\usepackage{amsthm}%
\usepackage{mathrsfs}%
\usepackage[title]{appendix}%
\usepackage{xcolor}%
\usepackage{textcomp}%
\usepackage{manyfoot}%
\usepackage{booktabs}%
\usepackage{algorithm}%
\usepackage{algorithmicx}%
\usepackage{algpseudocode}%
\usepackage{listings}%
\usepackage[pagewise]{lineno}

\theoremstyle{thmstyleone}%
\theoremstyle{thmstyletwo}%

\theoremstyle{thmstylethree}%

\begin{document}

\title[Article Title]{Anomalous Diffusion of Lithium-Anion Clusters in Ionic Liquids}

\author[1]{\fnm{YeongKyu} \sur{Lee}}

\author[2]{\fnm{JunBeom} \sur{Cho}}

\author[1]{\fnm{Junseong} \sur{Kim}}

\author*[2]{\fnm{Won Bo} \sur{Lee}}\email{wblee@snu.ac.kr}

\author*[1]{\fnm{YongSeok} \sur{Jho}}\email{ysjho@gnu.ac.kr}

\affil*[1]{\orgdiv{Department of Physics}, \orgname{Gyeongsang National University}, \orgaddress{\street{Jinjudae-ro 501}, \city{Jinju}, \postcode{52828}, \state{Gyeongsangnam-do}, \country{Rep. of KOREA}}}

\affil*[2]{\orgdiv{School of Chemical and Biological Engineering}, \orgname{Seoul National University}, \orgaddress{\street{1, Gwanak-ro}, \city{Gwanak-gu}, \postcode{08826}, \state{Seoul}, \country{Rep. of KOREA}}}


\abstract{Lithium-ion transport is significantly retarded in ionic liquids (ILs). In this work, we performed extensive molecular dynamics (MD) simulations to mimic the kinetics of lithium ions in ILs using [\emph{N}-methyl-\emph{N}-propylpyrrolidium (pyr$_{13}$)][bis(trifluoromethanesulfonyl)imide (Ntf$_{2}$)] with added LiNtf$_{2}$ salt. And we analyzed their transport, developing a two-state model and comparing it to the machine learning-identified states. The transport of lithium ions involves local shell exchanges of the Ntf$_{2}$ in the medium. We calculated train size distributions over various time scales. The train size distribution decays as a power law, representing non-Poissonian bursty shell exchanges. We analyzed the non-Poissonian processes of lithium ions transport as a two-state (soft and hard) model. We analytically calculated the transition probability of the two-state model, which fits well to the lifetime autocorrelation functions of LiNtf$_{2}$ shells. To identify two states, we introduced the graph neutral network incorporating local molecular structure. The results reveal that the shell-soft state mainly contributes to the transport of the lithium ions, and their contribution is more important in low temperatures. Hence, it is the key for enhanced lithium ion transport to increase the fraction of the shell-soft state.}


\maketitle
\section{Introduction}
Due to their high electrochemical and thermal stability, ionic liquids (ILs) are emerging as promising candidates for lithium battery solvents. Despite these advantages, the fact that ionic liquids are not electroactive poses a challenge to their application. Lithium ions are added as charge carriers, but anionic ILs surround them to form a solvation shell, which heavily slows down lithium-ion transport. So, for lithium ions to act as charge carriers, it is necessary to understand the dynamics of lithium ions in ionic liquids. 

Molinari \textit{et al.} showed that lithium ions have a negative transference number in a wide range of ionic liquids (ILs). This is because lithium ions can form clusters with anions due to the strong attraction between them. The ideal solution theory is not applicable in this case because it does not take into account the strong cation-anion correlations. Using concentrated solution theory, they showed that lithium-ion containing shells possess a net negative charge over a wide range of IL-based electrolytes~\cite{molinari1, molinari2}.

McEldrew and Goddwin \textit{et al.} also studied concentrated electrolyte systems using ILs. They showed that alkali cations can have negative effective charges at low mole fractions, leading to the formation of asymmetric ionic clusters and percolating ion networks. They introduced a novel thermodynamic model to describe reversible ion aggregation and gelation in concentrated electrolytes. This model can explain complex ion associations beyond simple ion pairing. It offers insights into ionic cluster populations, gel formation, and ion partitioning~\cite{McEldrew, McEldrew2, McEldrew3, McEldrew4}. Other subsequent work also confirms negative transference numbers because of strong interactions between cation and anions~\cite{subsequent}.

Introducing a polarizable model allows us to accurately describe static properties, but the enormous computational complexity prevents us from simulating the system long enough time to measure its dynamics~\cite{add_ref1, add_ref2}. For this reason, our understanding of the transport mechanism of lithium-anion clusters is still limited. The conventional concept describing lithium diffusion via vehicular and structural motion are very intuitive and useful~\cite{structural1,structural2,structural3,structural4,structural5,structural6,structural7}, but they are closer to the temporal motions. To describe the scaling behavior of the diffusive motion, we may need to define states that persists for a certain period of time. 

There are studies characterizing the structural motion of lithium ions under the ILs using a machine learning approach. Kahle \textit{et al.}~\cite{kahle} and Molinari \textit{et al.}~\cite{molinari3} characterized the structural properties of lithium ions by the machine learning aid trajectory analysis method in solid-state ionic conductors. They mapped system coordinates into a specified space called a landmark basis to learn and encode correlation structures. By doing this, computation cost of conductivity was remarkably reduced, and a systematic reduction of uncertainty in conductivity was possible.

The Google Deepmind group recently reported that graph-neutral networks (GNNs) might classify the kinetical states of glassy particles into two states. They classified the two states of the glassy particles based on the frequency of neighbor exchange. A particle is classified as "hard" if it is unlikely to exchange its neighbors and as "soft" if it is likely to do so. The energy barrier for an exchange event is higher and stiffer in the "hard" state. They showed that GNNs successfully identified the soft particles that easily exchange their neighbors in the cage and hard particles that tend to move together with their neighbors in the cage. This  machine-learning aid classification resembles the idea of vehicular and structural motions but is more quantitative~\cite{gnn}. GNN offers the benefit of predicting the kinetic states of lithium ions with high accuracy. It connects the local structure with the kinetic behaviors. However, the criteria of state classification are not intuitive.

In this study, we find that the shell exchange of the lithium-ion solvation shell is bursty, which a single Arrhenius mechanism cannot explain. As a minimal approximated model, we develop a two-state model to explain the anomalous dynamics inspired by the conventional lithium-ion diffusion classification. We fit the autocorrelation function from the simulation to our model with two states identified by machine learning.  

\section{Simulation Details}
To properly implement the dynamics of ILs, a polarizable force-field ought to be used. The non-polarizable force-field of ILs underestimates their dynamics~\cite{nonpolar1, nonpolar2, nonpolar3, nonpolar4}. For instance, the non-polarizable force-field simulation of [Li][Ntf$_{2}$] mixed with 1,3-dioxolane and 1,2-dimethoxyethane yields diffusion coefficients $100$ times smaller than the experimental results. Thus, incorporating polarization is essential to predict and understand the lithium-ion transport in the ILs medium. APPLE\&P is known to be a reliable force-field, developed by Borodin \textit{et al.}~\cite{first_md}. It includes many-body polarization effects and calculates induced dipole interactions for all atoms in the system. With the inclusion of polarization effects, vaporization enthalpy, and other physical properties are closely estimated to the experimental values~\cite{first_md}.

We prepared the [PYR$_{13}$]$^{+}$[Ntf$_2$]$^{-}$ as electrolyte doped with $5\%$ of LiNtf$_2$ compared to [PYR$_{13}$][Ntf$_{2}$]. Initial configurations were generated by PACKMOL~\cite{packmol}. We randomly inserted them into a cubic box and made outputs in XYZ format. After inserting the molecules, we modified the .xyz format into .cc1 format. The modified .cc1 files were processed with a system generator in the WMI-MD simulation package to assign APPLE\&P force-fields to the molecules in the system~\cite{Borodin2009}. We adopted a multiple-time step integrator developed by Martyna et al.~\cite{multistep}. Multiple time steps integrate interactions in three different time scales. The shortest time step of $0.5$ fs is for bonds, bends, and improper torsions. The intermediate step of $1$ fs is for torsions, short-ranged nonbonded interactions within a $7$ {\AA}  truncation. And the longest timestep of $2$ fs is for nonbonded interactions between $7$ {\AA}  and $11$ {\AA} as well as for the calculation of Coulomb interactions with Particle Mesh Ewald (PME) summation. Simulations were done at $298$ K, $353$ K, $373$ K, and $423$ K. 

We first equilibrated initial configurations at $500$ K for $100$ ps under the NPT ensemble and then ran an additional $1$ ns at the target temperatures. The production run was conducted for $20$ ns for $353$ K, $373$ K, $423$ K, and $100$ ns for $298$ K in the NVT ensemble at the target temperatures. All simulations were performed using WMI-MD simulation packages.
\section{Results and Discussion}
RTIL consists of ions that are in a liquid phase at room temperature. The strong Coulombic interaction among ions causes the formation of ion-pairs and more complex clusters that significantly slow the diffusion of ions in the ionic liquid~\cite{McEldrew}. The conventional electrostatic theories, which treat ions as single charges, are invalid for ionic liquids. The frustration in the room temperature critically hinders the transport mechanism. Most previous simulations were performed at high temperatures and extrapolated the high-temperature results to room temperature~\cite{add_ref3, add_ref4, add_ref5}. Although it is too computationally expensive to sample the whole dynamics ergodically, it is still tractable to sample local cluster kinetics within a reasonable time. We attempt to rationalize the mechanism using the kinetics of lithium-anion clusters and connect it to their transport. 

Fig.~\ref{fig:se} plots the exchange events of the solvation shell members over time. Interestingly, the shell exchange occurs not homogeneously but bursty. The shell exchanges are more concentrated in some small segments than in the rest part. Later we will show that the process is indeed non-Poissonian. The heterogeneous dynamics of bursty behavior come from the collective behavior, such as long-range correlation, memories, or multiple states effect~\cite{karsai2018bursty}. Among the possibilities, the two-state idea is consistent with the vehicular and structural motion classification.

Under the addition of extra Li$^{+}$, Li$^{+}$ brings anions nearby to form a solvation shell. Li$^{+}$ and anions at the solvation shell move together as a single entity with a larger mass. This fact tempts us to separate the motion of Li$^{+}$ as a vehicular motion like part, which preserves the members of the single entity during movement, and a structural motion like part, in which Li$^{+}$ hops by exchanging one of Ntf$_{2}^{-}$ in the solvation shell.
However, separating the vehicular motion from the structural motion is a bit obscure because Li$^{+}$ movements always accompany both. Here we propose a more straightforward classification for diffusion. 

We classified the state of the clusters from explicit events of anion shell exchange. The transition rate of the anions in the shell is proportional to
\begin{equation}
    \tau^{-1} \propto \exp\left\{ - \frac{\Delta U(s)}{k_B T}\right\}
    \label{eq:arrhen}
\end{equation}
where $\Delta U(s)$ is the shell exchange energy barrier at configurational states $s$. If only a single state exists, and the exchange event occurs randomly by thermal fluctuation (Poissonian process), the transition path will be dominated by the single most efficient one from the Kramer's transition theory.

We generated time series tracks of the change in the solvation shell composition, recording all the events that anions cross the border, as shown in Fig.~\ref{fig:se}. The cut-off distance of the border was taken from the first minimum of the radial distribution function of Li$^+$-Ntf$_{2}^{-}$. The distance was measured between the centers of masses of the ions.  

\begin{figure}
    \centering
    \includegraphics[width=0.5\textwidth]{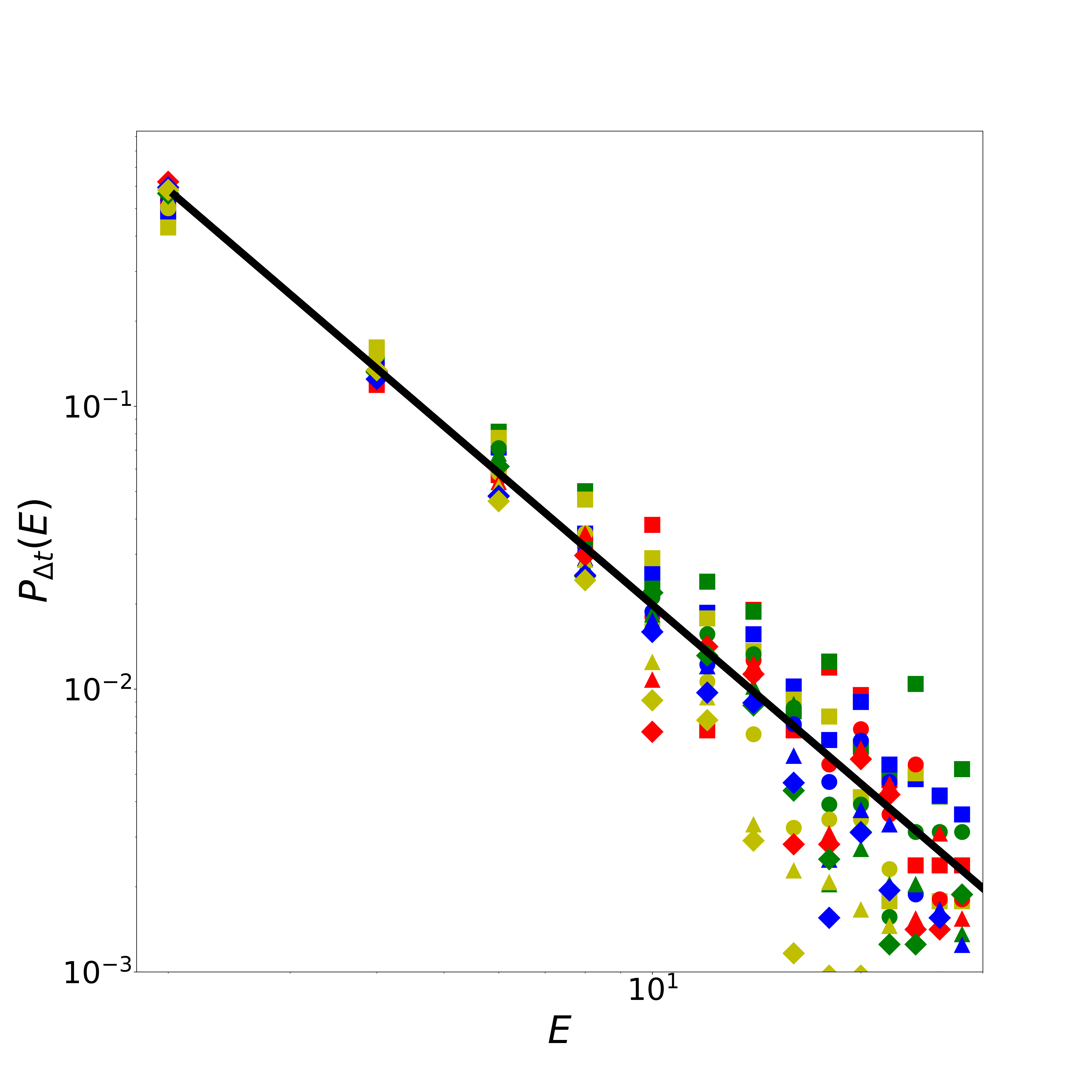}
    \caption{Probability distribution functions of the number of transition $E$ in the train size $\Delta t$, where E refers to the number that anions cross border}; {\footnotesize red, green, blue, and yellow color denotes 298 K, 353 K, 373 K, and 423 K, respectively. Square, triangle, circle, and diamond symbols indicate $\Delta t$, $\Delta t/2$, $\Delta t/3$, and $\Delta t/4$, respectively.}
    \label{fig:f1}
\end{figure}
We calculated the bursty train size distribution of the sequential shell exchanges. The events are in the same train if consecutive events occur within $\Delta t $. We defined the train size $\Delta t$ as a function of the average inter-event time, and varied it by the fractions of $1/1$, $1/2$, $1/3$, and $1/4$. The bursty train size distribution in $\Delta t$ follows a power law, $P_{\Delta t}(E) \propto E^{-\beta}$ indicating that the system is bursty. We found that $ \Delta t $ does not change the power law exponent. 

Fig.~\ref{fig:f1} plots the burst size distribution as a function of train sizes. The distribution decays with the same exponent $\beta=1.3$ for the different $\Delta t$, indicating the bursty behavior, which usually is from a long-range correlation, memory effect, or the existence of multiple states~\cite{karsai2018bursty}. Recently, Feng \textit{et al.} showed that a two-state model might be suitable for the kinetics of binary ionic liquid mixtures~\cite{twostatemodel}.

Fig. S1 and Fig. 1 suggest that a two-state model might cause the bursty behavior. We develop a formula for trajectory averaged quantities between two states using a discrete Markov chain Monte Carlo model to explain the underlying mechanisms of lithium transport. 

We discretized $t$ as $n\delta t$, where $n$ is the number of steps and $\delta t$ is the unit time for a single step. The survival probability $\xi_i(n)$ refers to the probability that a particle stayed in the state $i$ remains in the same state during $n$ steps without undergoing transitions of states. We assume that it transits to the other state randomly,
\begin{equation}
    \xi_i(n) = \exp\left(-\alpha_i n\right),
\end{equation}
where $\alpha_{i}$ indicates the transition probability at state $i$, and $i$ denotes $1$ or $2$. 
We can change it to the discrete probability $\xi_i(n) = \left(1-\alpha_{i}\right)^n$, which only changes some constant factors in the final results. 

Then, the transition rate $h_i(n)$ is
\begin{equation}
    h_i(n) = -\frac{\Delta \xi_{i}(n)}{\Delta n} = (1-\exp\left(-\alpha_{i}\right))\xi_{i}(n).
\end{equation}
$h_i(n) \Delta n$ is the probability that the system stays in the state $i$ during $(0,n)$ and then undergoes a transition in $(n,n+1)$. 

We assumed that the two-state model is a mixture of two Poisson processes. We defined Li$^+$ is in state $s_{1}$ if the dissociation of Li-Ntf$_2$ pairs is rare, otherwise, $s_{2}$. 
Thus, the lifetime of Li-Ntf$_2$ pairs would be shorter at $s_{2}$. The dissociation rate of Li-Ntf$_{2}$ pairs in the state $i$ can be expressed as survival probability functions
\begin{eqnarray}
    f_{i}(n) = \exp\left(-\beta_{i} n\right) \nonumber
\end{eqnarray}
where $\beta_{i}$ is an inverse of the average survival time of Li-Ntf$_{2}$ pairs.
\begin{figure}
    \centering
    \includegraphics[width=0.5\textwidth]{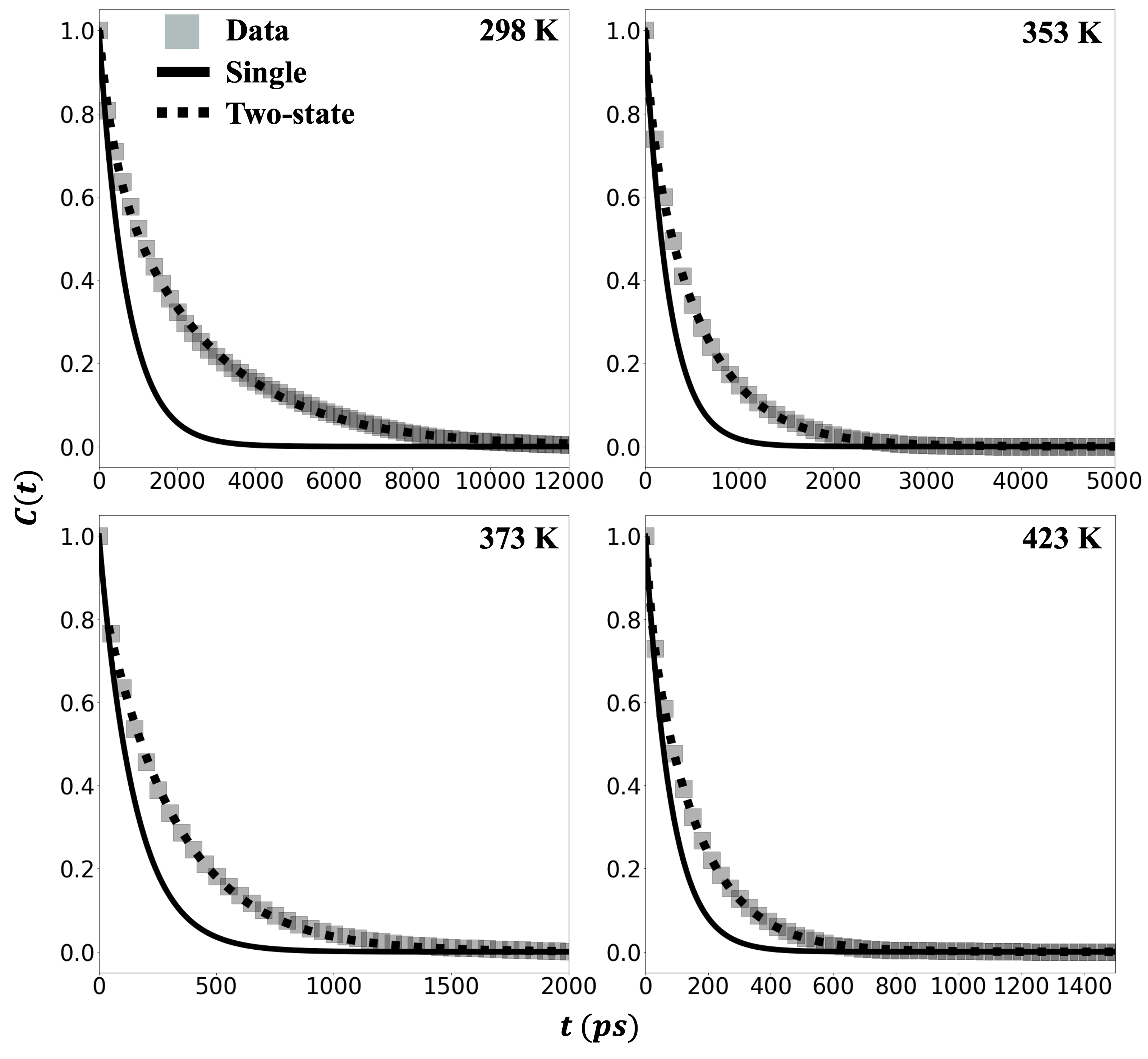}    
    \caption{ACFs (black square) fitted into single (solid) and two-state model analytical function (dotted); \footnotesize{residence times are 19.329 ns (298 K), 4.926 ns (353 K), 2.885 ns(373 K), 1.406 ns (423 K)}. log scaled ACFs compared with Kohlrausch–Williams–Watts (KWW) function are provied in Fig.~\ref{fig:log_log}.}
    \label{fig:f2}
\end{figure}
We define $s_{i\bar{i}}(n)$ the average survival probability Li-Ntf$_{2}$ pairs going from state $i$ at time $0$ to state $\bar{i}$ at time $n$. Other trajectory-averaged quantities can be calculated similarly. Then, we found $s_{i\bar{i}}(n)$ are expressed as simultaneous recurrence relations,
\begin{equation}
\begin{split}
s_{ii}(n) &= \sum\limits_{k=0}^{n}h_{i}(k)f_{i}(k)x_{\bar{i}i}(n-k) + \sum\limits_{k=n}^{\infty}h_{i}(k)f_{i}(n) \\
s_{\bar{i}i}(n) &= \sum\limits_{k=0}^{n}h_{\bar{i}}(k)f_{\bar{i}}(k)x_{ii}(n-k) - h_{\bar{i}}(0)\delta_{n,0},
\label{eq:sij}
\end{split}
\end{equation}
where $i \neq \bar{i}$, \textit{i.e.} if $i=1$, then $\bar{i}=2$, and vice versa. $s_{ii}(0) = 1$ and $s_{\bar{i}i}(0) = 0$. 
The calculation of the simultaneous equations is tractable under the Z-transformation.  The average of Li-Ntf$_{2}$ pair dissociation over two states is then
\begin{equation}
\begin{split}
    s_{tot}(n) &= r_{1}\left(s_{11}(n) + s_{12}(n)\right) \\
               &+ r_{2}\left(s_{21}(n) + s_{22}(n)\right),
\end{split}
\end{equation}
where $r_{i}$ is the fraction of each state (we will obtain the fraction from the GNNs later).

To verify our two-state model, we fitted the pair autocorrelation function (ACFs) of Li-Ntf$_2$ with the two-state model. The ACFs are calculated by 
\begin{equation}
    c(t) = \frac{1}{T-t}\int_{0}^{T-t} \frac{1}{N_{t'}} \sum_{i=1}^{N_{t'}} \psi_{i}(t'+t)\psi_{i}(t') dt'.
\end{equation}
$N_t'$ is the number of Li-Ntf$_2$ pairs at time $t'$. $\psi(t)$ is unity if Li-Ntf$_2$ pair survives for time $t$. If not, it is zero. The ACFs are averaged over all the Li-Ntf$_2$ pairs during the whole simulation trajectories.   

Fig.~\ref{fig:f2} exhibits the ACFs with comparisons to a single exponential and two-state kinetics model. A single exponential model only fits a short-time scale. Our two-state model works well with ACFs for the whole time scale. Since our model considers both the shells and the state transitions over time $n$ in each state $i$, we can explain the two-state kinetics of lithium ions inside the ILs.
\begin{figure}
    \centering
    \includegraphics[width=0.5\textwidth]{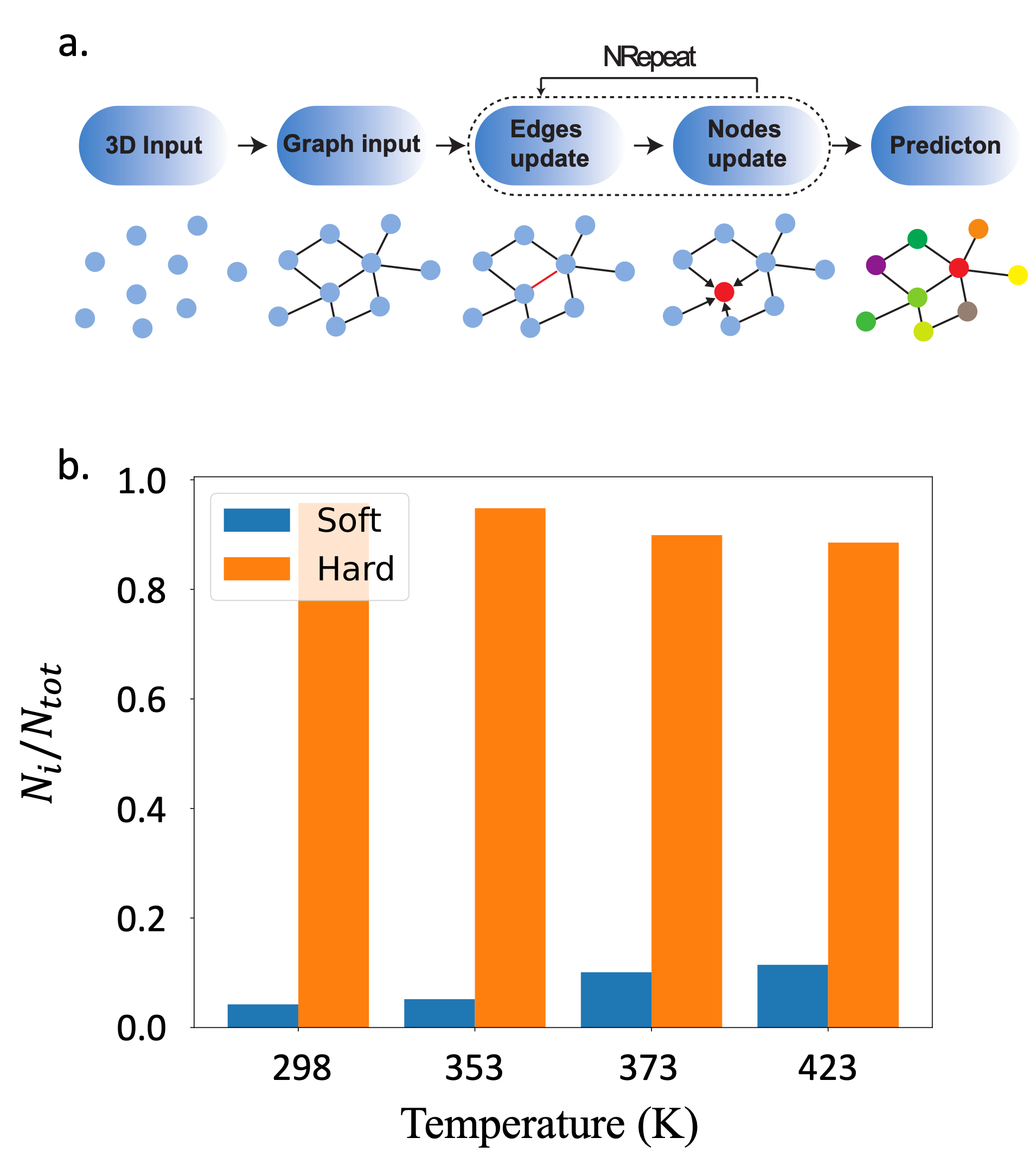}
    \caption{Schematic workflow of the GNNs and results; a. Schematic algorithm of GNNs, b.Ratio of the soft and hard particles; N$_{i}$ denotes the states ($i=$ soft, hard)}
    \label{fig:fig3}
\end{figure}

To consider the molecular level features of the two states, we applied GNNs for the two-state classification, which successfully identified two states of glass~\cite{gnn, github, web}. Here are some details of the GNNs. We first labeled the lithium ions depending on the train size. If the number of transitions exceeds $E_{c}$ within train size $\Delta t$, we labeled them soft-state. Otherwise, we labeled them hard-state. Next, we mapped 3D configurations of whole system obtained from the molecular dynamics simulations to graph input. We treat the centers of mass of every molecule as nodes. We connected each node if two adjacent nodes were within $r_{cut}$. The cutoff is taken from the first shell minimum of the RDFs. The RDFs is presented in section A.3 of Supporting Information. We embedded initial node values depending on the types of molecules, 0, 1, and 2, for the lithium, Ntf$_2$, and PYR, respectively. We also embedded initial edge values as relative positions between nodes. 

We update edge and node features using the previously embedded values of the edges and nodes. We update edge and node features of every time step. We repeated these steps up to seven times~\cite{gnn}, \textit{i.e.} information is mixed over seven layers as depicted in Fig.~\ref{fig:fig3}.a. It turns out that there were no significant differences in prediction results over the repetitions in edge and node updates which means the first few shells are enough to characterize the Li$^+$-Ntf$_{2}^{-}$ complex. It is plausible that the strong coordination between Li$^+$-Ntf$_{2}^{-}$ makes the first shell persistent.  Finally, we made predictions on the lithium ions based on the labels we assigned above.

We varied train size $\Delta t$ to find optimal $\Delta t$ and E values, yielding the number of transitions for each temperature. We calculated accuracy, precision, recall, and F1-scores (presented in section 4 of Supporting Information). Optimal $\Delta t$ and E are obtained at maximum F1-scores. GNN results predicted by the optimal $\Delta t$ and $E$ are presented in Fig.~\ref{fig:fig3}.b. Indeed, there is some connection between local shell structures and the shell exchange kinetics, and GNN can successfully identify it. 

\begin{figure}
    \centering
    \includegraphics[width=0.5\textwidth]{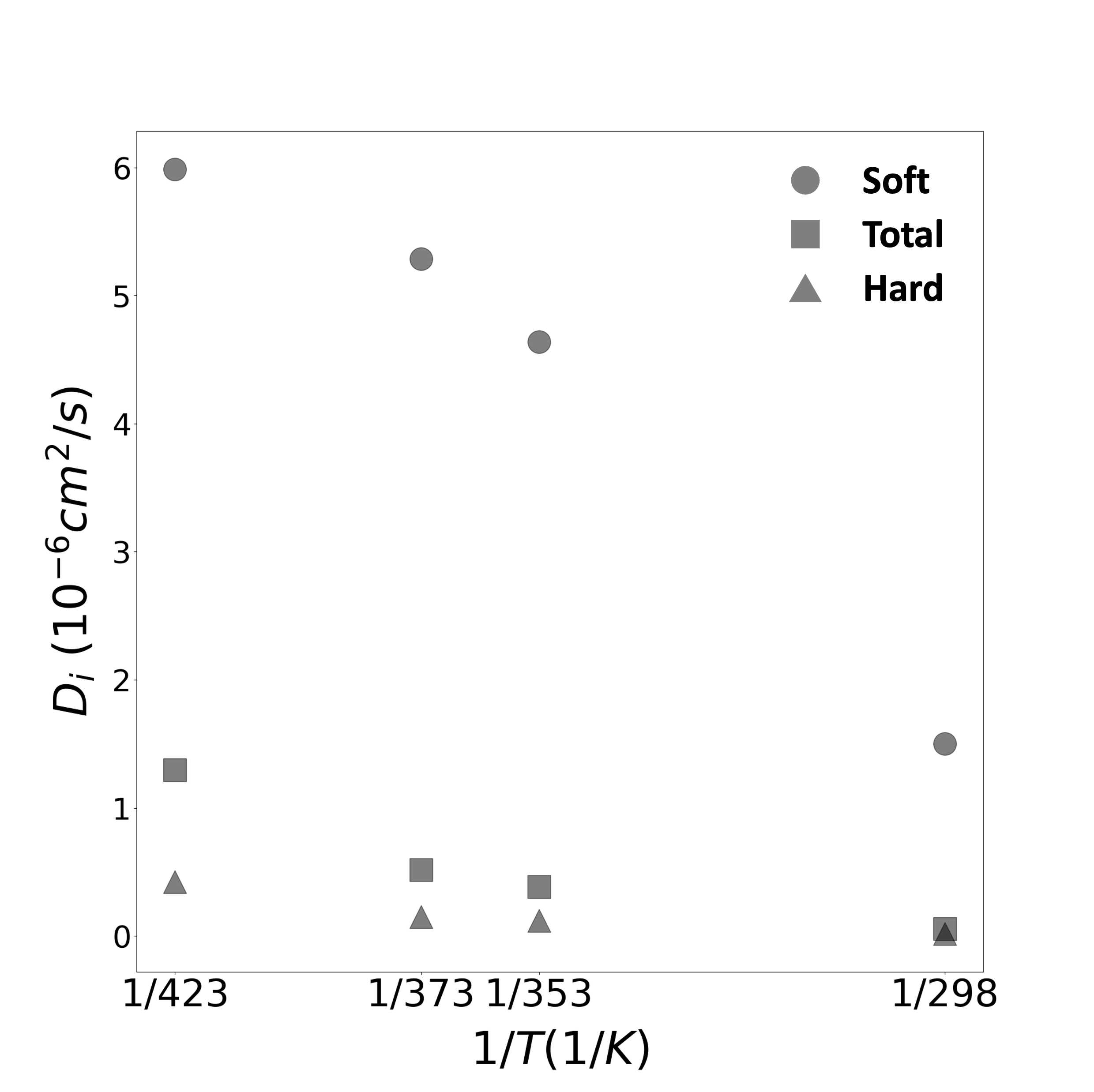}
    \caption{Diffusion coefficients of soft state, hard state, and the overall average}
\end{figure}

We calculated the diffusion coefficients of lithium ions at two states for various temperatures. The diffusion is very low at room temperature. 
The mean-squared displacement (MSD) for lithium ions is,
\begin{equation}
	\begin{split}
		MSD = \frac{1}{N} \sum_{i=1}^{N} \left(\textbf{x}_{i}(t) - \textbf{x}_{i}(0)\right)^{2} = 6Dt.
	\end{split}
\end{equation}

It is hard to calculate the diffusion coefficients of the soft states because of their short residence time. Instead, we first calculated the diffusion coefficients of the hard state and over the whole trajectory. Then, we can estimate the diffusion coefficients of the soft state from them because the total diffusion coefficient is a weighted sum of the diffusion coefficient of hard and soft states. The weights are obtained from the GNN classification of hard and soft states. This holds as long as the two diffusion processes are independent. 

Through the temperatures, the diffusion coefficient of the soft state is several times larger than that of the hard state. 
The absolute values of the diffusion coefficient are smaller at low temperatures, as expected. It is plausible that the kinetics is slow due to the constraint inhibiting shell exchanges. But the diffusion of the soft state is sufficiently large even at room temperature.
An interesting difference in diffusion over the temperatures is the contribution of the soft state Li$^+$ in the overall diffusion. It shows moderate fractions of the soft state in high temperatures, whereas the fraction is negligible at room temperature. It is why the diffusion at room temperature decreases substantially despite the diffusion of the soft state being fast enough even at room temperature. Thus, increasing the fraction of the soft state would be critical to accelerating the Li$^{+}$ transport. In terms of qualitative arguments, Li$^{+}$ diffuses via vehicular motion at room temperature. We need to increase the fraction of structural diffusion to enhance its transport.

As previously proposed by several researchers~\cite{structural1, structural2, structural3, structural4, structural5, structural6, structural7}, the kinetics of lithium-ion transport can potentially be understood through both vehicular and structural motion. However, these mechanisms mainly capture localized structural properties and short-term kinetics of lithium-ion transport, which may not persist long enough to capture the scaling behavior of mean square displacement. Thus, our two-state model may compensate for the intuitive explanation for structural and vehicular motions. However, there is a limitation in understanding the physical meaning of $E_{c}$ which is optimized by the machine learning tool.

\section{conclusion}
We found that the shell exchange kinetics of Li$^{+}$ is bursty. Most shell exchanges are concentrated in some small time windows. We developed a two-state model to explain the bursty behavior assuming that the exchange events are Poissonian in each state while the transitions between two states are Markov. We presented a formula for trajectory averaged quantities. The two states are classified using GNN learning, which successfully connects the local solvation shell structure to the exchange kinetics.

Although the conventional classification of diffusive motions, structural diffusion and vehicular diffusion, are highly intuitive and useful to the understanding of the underlying mechanism, they describe a temporal motion of the particles. To obtain the scaling behavior for diffusive motion, we need a better judgment of the states which persists for a long time. Our approach reinforces the classical classification of two diffusive motions by complementing the weakness of the classification.

The significant drop in ionic transport at low temperatures is due to the reduced soft state in which the shell exchanges are frequent. As shown in section A.6 of Supporting Information, the exchange rate in the soft state is less sensitive to temperatures than in the hard state. Thus, increasing the soft state's fraction would improve the lithium-ion transport at room temperature.

Often adding organic solvents and introducing a ternary ILs system is an excellent strategy to reduce the viscosity of the medium. Adding organic solvents or molecules that weakly interact with lithium ions or anions can loosen the lithium-ion clusters. It will lead their kinetic state to another level, such as a soft state or even a softer state, which eventually enhances ion transport. 
\bibliography{sn-article}
\clearpage
\begin{appendices}
\newpage
\renewcommand\thefigure{A\arabic{figure}}
\section{Supporting Information}
\setcounter{figure}{0}
\renewcommand\thefigure{A\arabic{figure}}

\subsection{Shell Exchange Analysis}
\begin{figure}[htbp]
    \centering
    \includegraphics[width=0.5\textwidth]{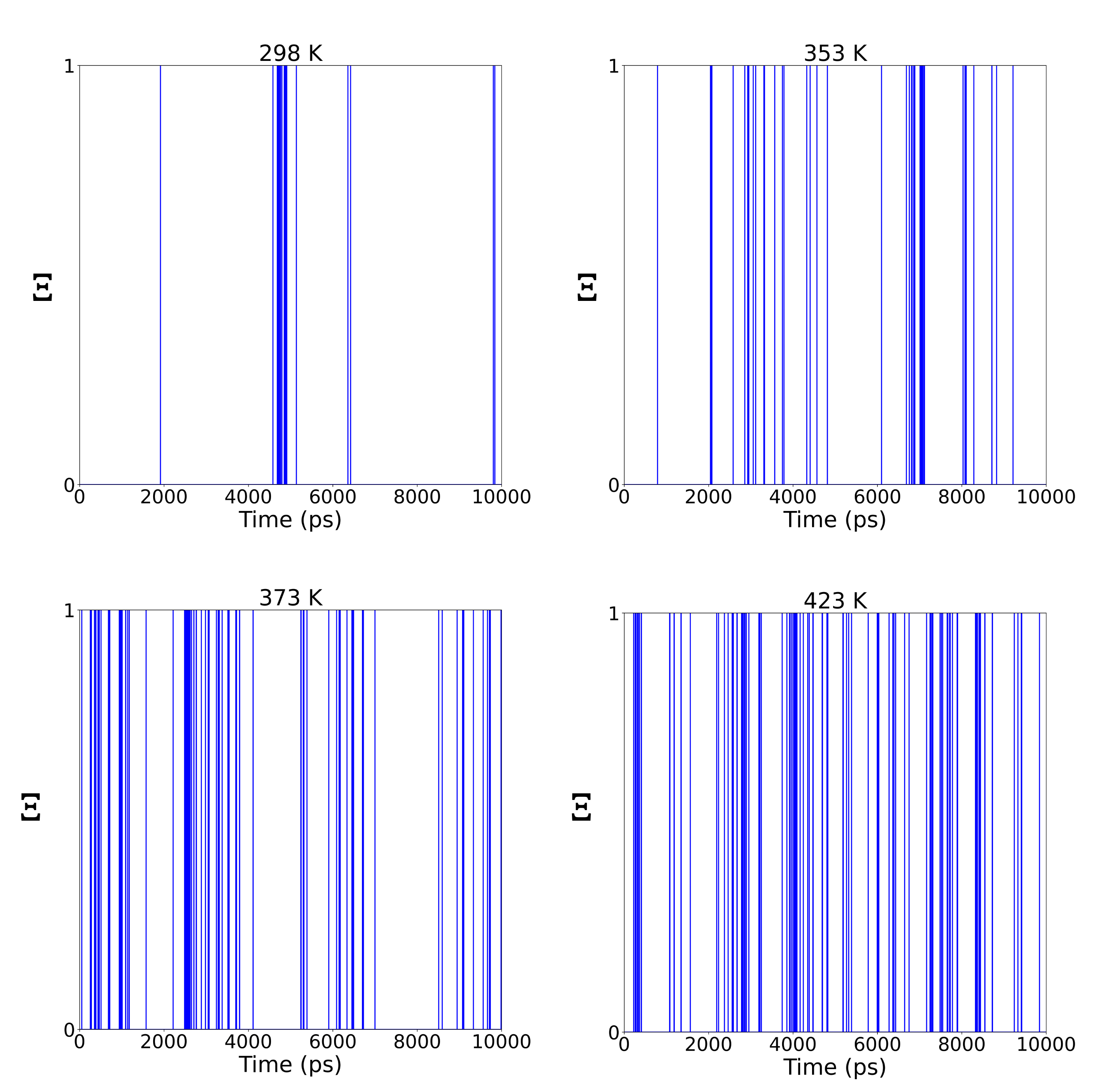}
    \caption{Shell exchanges at different temperatures}
    \label{fig:se}
\end{figure}

Shell exchange is calculated as follows:
\begin{equation}
    \Xi =
    \begin{cases}
        1, & \text{for}\,\,\,\,s_{t-1} \neq s_{t} \\
        0, & \text{for}\,\,\,\,s_{t-1} = s_{t}
    \end{cases}
    \label{eq:timeseries}
\end{equation}
.
\subsection{Two-state kinetics model}
We first discretize continuous time $t$ into the number of steps $n$. We define survival function as
\begin{equation}
    \xi_{i}(n)=e^{-\alpha_{i}n},    
\end{equation}
which yields the probability that a particle in a state $i$ at initial remains in the same state $i$ during $n$ steps without undergoing any transitions. Here, $i$ can be 1 or 2. 
We can change it to the discrete probability $\left(1-\alpha_{i}\right)^n$, which only changes some constant factors in the final results. 

If we define $h_{i}(n)$ as a time derivative of survival function $\xi_{i}(n)$,
\begin{equation}
    h_{i}(n) = - \frac{\Delta \xi_{i}(n)}{\Delta n} = \left(1-e^{-\alpha_{i}}\right)\xi_{i}(n),
\end{equation}
we obtain the probability that the system stays in the state $i$ for time $n$ and then undergoes a transition in $(n, n+\Delta n)$ as $h_{i}(n)\Delta n$. 

We also define the probability that Li$^{+}$ in the state $i$ stays in the same state for $n$ steps, 
\begin{equation}
    f_{i}(n) = e^{-\beta_{i} n}.
\end{equation}

We will calculate Li-Ntf$_{2}$ pair dissociation over two states during time $n$. $s_{11}(n)$ is the survival probability of Li-Ntf$_{2}$ pair which was in the state $1$ at step $0$ and ended up with in the state $1$ at step $n$. We don't care about transitions in the intermediate steps. We can apply this measure for any trajectory-averaged quantities. 
\begin{equation}
\begin{split}
    s_{11}(n) &= \sum\limits_{k=0}^{n} h_{1}(k)f_{1}(k)x_{21}(n-k) + \sum\limits_{k=n}^{\infty}h_{1}(k)f_{1}(n)
\end{split}
\end{equation}
where $s_{11}(0)=1$.
The last term is the sum of the probabilities that Li-Ntf$_{2}$ pair undergoes transition after $n$ steps, \textit{i.e.} it corresponds to the probability the pair do not undergo any transition during $n$ steps. 
In a similar way, $s_{21}(n)$ (average survival probability of Li-Ntf$_{2}$ pairs which started in state $2$ at initial and ended up in state $1$ at final.) is expressed as a function of $s_{11}(n)$,
\begin{equation}
\begin{split}
    s_{21}(n) &= \sum\limits_{k=0}^{n} h_{2}(k)f_{2}(k)s_{11}(n-k) - h_{2}(0)\delta_{n,0},
\end{split}
\end{equation}
where $s_{21}(0) = 0$. 
The simultaneous recurrence relation is solved under the Z-transformation,
\begin{equation}
\begin{split}
    s_{11}(z) &= h_{1}(0)G_{1}(z)s_{21}(z) + G_{1}(z) \\
    s_{21}(z) &= h_{2}(0)G_{2}(z)s_{11}(z) - h_{2}(0).
\end{split}
\end{equation}
Using $g_{i}(n) = \xi_{i}(n)f_{i}(n)$, where $G_{i}(z)$ is a Z-transform of $g_{i}(n)$, we obtain,
\begin{equation}
\begin{split}
    s_{11}(z) = \frac{G_{1}(z)\left(1-h_{1}(0)h_{2}(0)\right)}{1-h_{1}(0)h_{2}(0)G_{1}(z)G_{2}(z)}.
\end{split}\label{eq:s11zt}
\end{equation}
Finally, the inverse z-transformation provides $x_{11}(n)$
\begin{equation}
\begin{split}
    s_{11}(n) &= \left(1-h_{1}(0)h_{2}(0)\right) \\
    &\times\left(\frac{z_{1}^{n}(z_{1}-\textbf{e}_{2})}{z_{1}-z_{2}} + \frac{z_{2}^{n}(\textbf{e}_{2}-z_{2})}{z_{1}-z_{2}}\right),
\end{split}
\end{equation}
where $\textbf{e}_{i} = e^{-\alpha_{i}-\beta_{i}}$, and $z_{1}$ and $z_{2}$ are the solutions of the quadratic equation which is appealed in the denominator of Eq.~\ref{eq:s11zt}. 

Following the same procedures, we can find $s_{22}(n), s_{12}(n)$ and $s_{21}(n)$ as 
\begin{equation}
\begin{split}
s_{12}(n) &= h_{1}(0)\left(-\frac{e_{1}e_{2}}{z_{1}z_{2}}\delta_{n,0} \right. \\
&\left. + z_{2}^{n}\frac{e_{1}e_{2}-e_{1}z_{2}-e_{2}z_{2}}{z_{2}\left(z_{1}-z_{2}\right)} \right. \\
&\left. + z_{1}^{n}\frac{e_{1}z_{1}+e_{2}z_{1}-e_{1}e_{2}}{z_{1}\left(z_{1}-z_{2}\right)}\right) \\
s_{21}(n) &= h_{2}(0)\left(-\frac{e_{1}e_{2}}{z_{1}z_{2}}\delta_{n,0} \right. \\
&\left. + z_{2}^{n}\frac{e_{1}e_{2}-e_{1}z_{2}-e_{2}z_{2}}{z_{2}\left(z_{1}-z_{2}\right)} \right. \\
&\left. + z_{1}^{n}\frac{e_{1}z_{1}+e_{2}z_{1}-e_{1}e_{2}}{z_{1}\left(z_{1}-z_{2}\right)}\right) \\
s_{22}(n) &= \left(1-h_{1}(0)h_{2}(0)\right) \\
    &\times\left(\frac{z_{1}^{n}(z_{1}-\textbf{e}_{1})}{z_{1}-z_{2}} + \frac{z_{2}^{n}(\textbf{e}_{1}-z_{2})}{z_{1}-z_{2}}\right)
\end{split}
\end{equation}  
The average of Li-Ntf$_{2}$ dissociation over two states is then represented as
\begin{equation}
\begin{split}
s_{tot}(n) &= r_{1}\left(s_{11}(n) + s_{12}(n)\right) \\
           &+ r_{2}\left(s_{21}(n) + s_{22}(n)\right).
\end{split}
\end{equation}

We would like to remark that we can obtain the same formula starting from the discrete probability functions $\xi_{i}(n)=\left(1-\alpha_i\right)^n$ except for some constant factors. 

\newpage
\subsection{Radial Distribution Functions}
\begin{figure}[htbp]
    \centering
    \begin{subfigure}[b]{0.4\textwidth}  
        \centering
        \includegraphics[width=\textwidth]{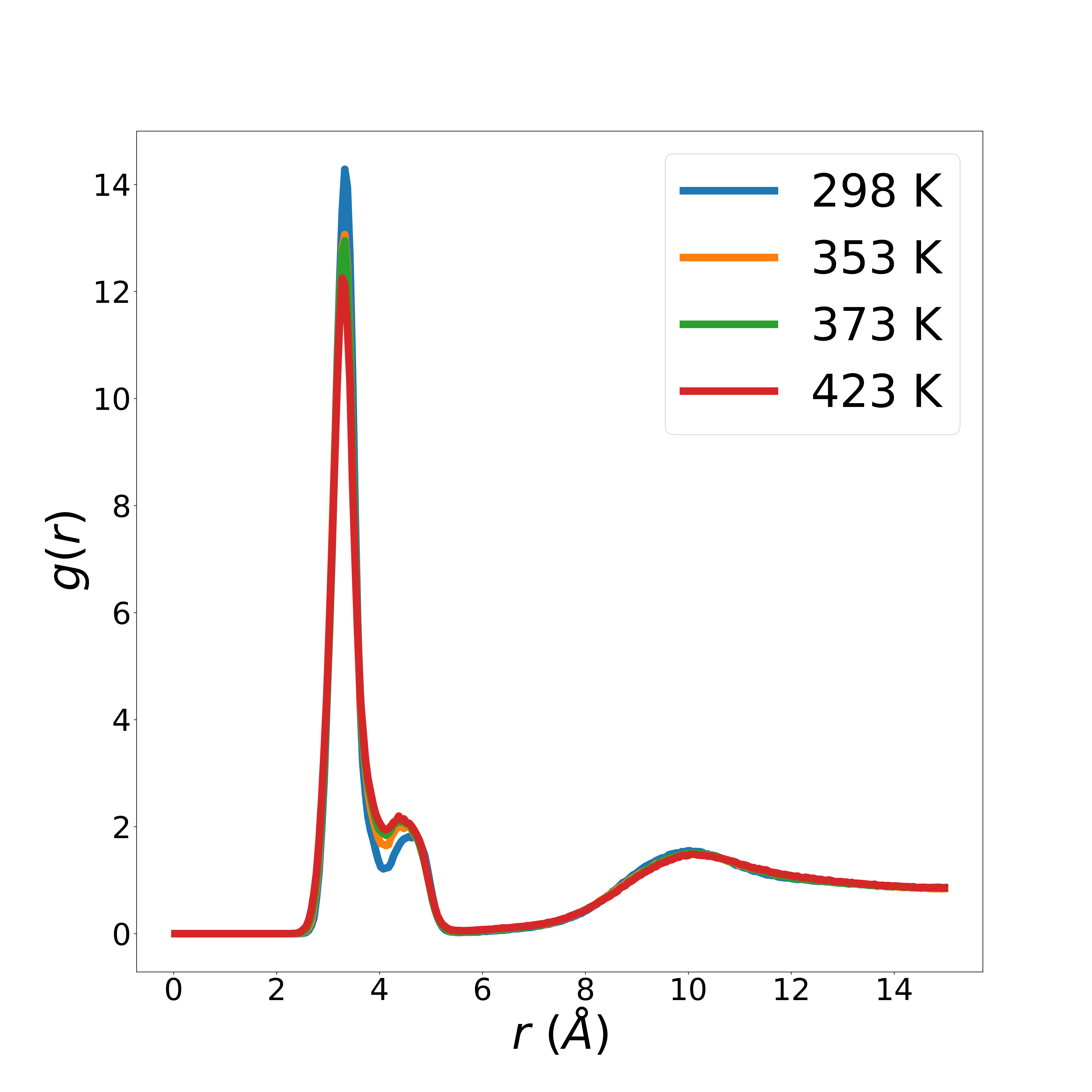}
        \subcaption{RDF of Li-Ntf$_{2}$}
    \end{subfigure}
    \hfill
    \begin{subfigure}[b]{0.4\textwidth}   
        \centering 
        \includegraphics[width=\textwidth]{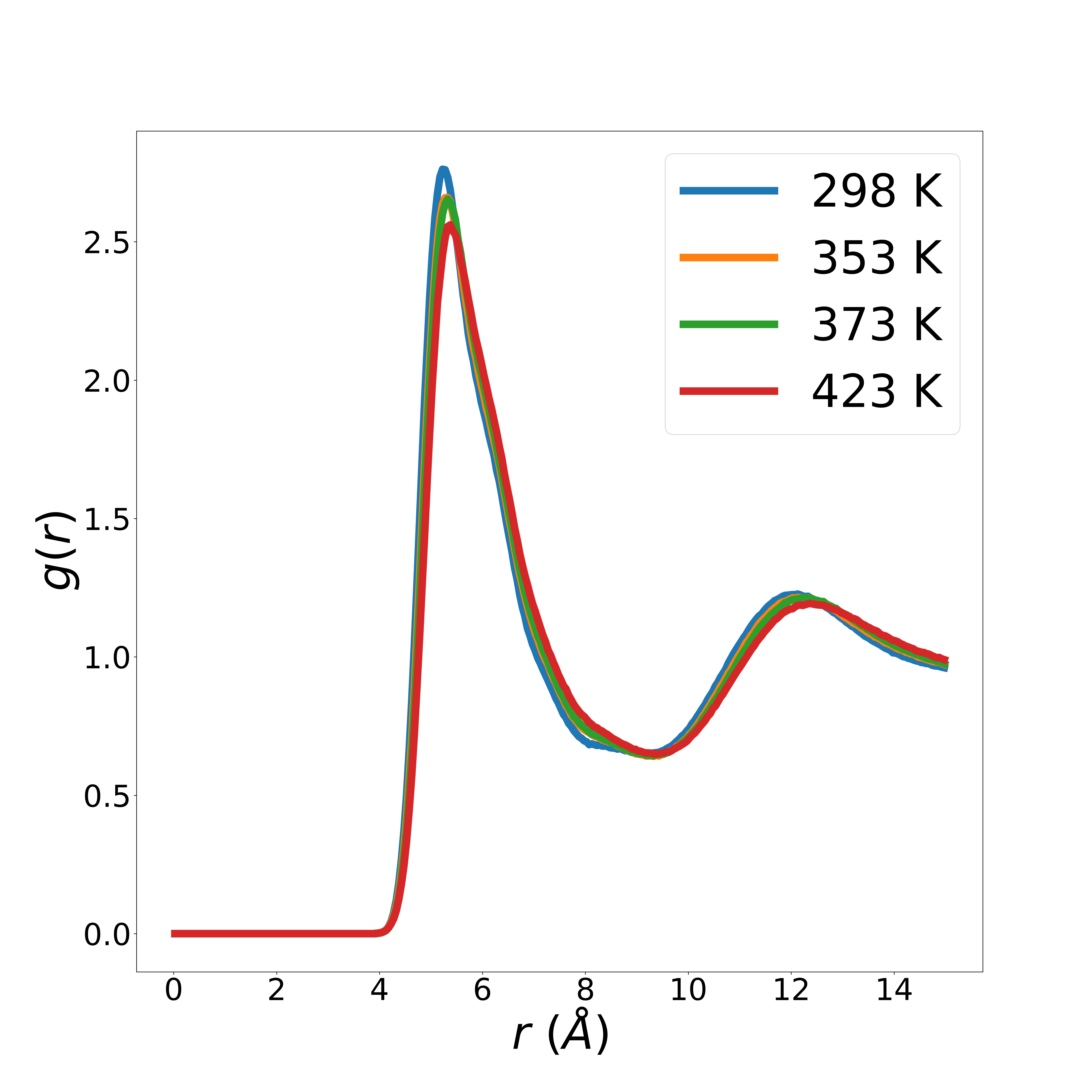}
        \subcaption{RDF of PYR-Ntf$_{2}$}
    \end{subfigure}
    \caption{Radial distribution functions calculated from the molecular dynamics simulation trajectory; we connected Li-Ntf$_{2}$ if the distance is within 5.2 \AA, and also conntected PYR-Ntf$_{2}$ if the distance is less than 9 \AA}
    \label{table:rdf}
\end{figure}
\clearpage
\subsection{Results of Accuracy, F1-Score, Recall, and Precision}
We calculated accuracy, F1-score, recall, and precision from the confusion matrix.
Precision is the accuracy of positive predictions, recall measures the ability to capture all positive instances, F1-score combines precision and recall, and accuracy measures overall correctness in a classification task. Each of these metrics provides the performance of a classification model and the choice of the most appropriate metric.
\begin{figure}[htbp]
    \centering
    \includegraphics[width=0.5\textwidth]{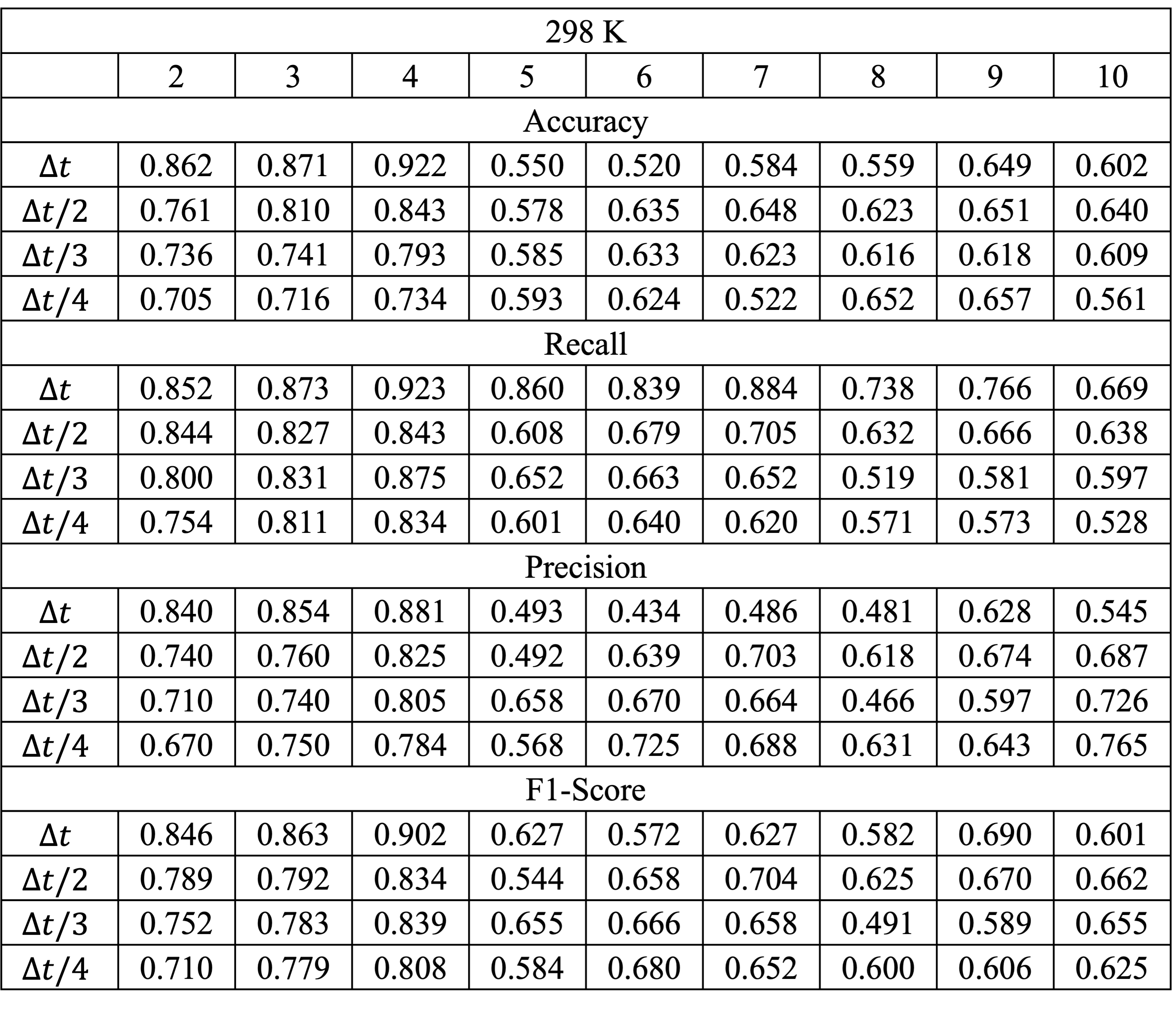}
\end{figure}
\begin{figure}[htbp]
    \centering
    \includegraphics[width=0.5\textwidth]{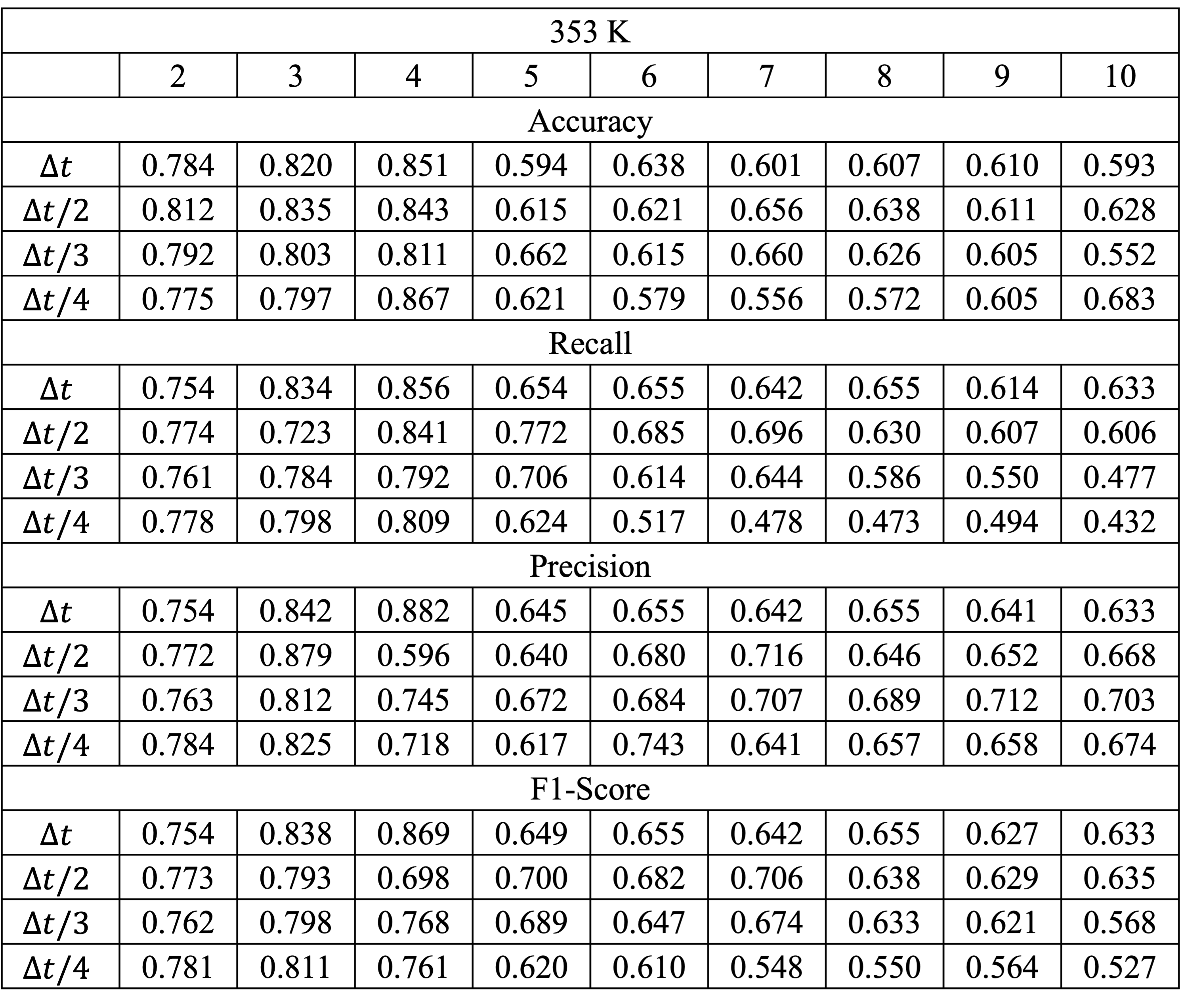}
\end{figure}
\newpage
\begin{figure}[htbp]
    \centering
    \includegraphics[width=0.5\textwidth]{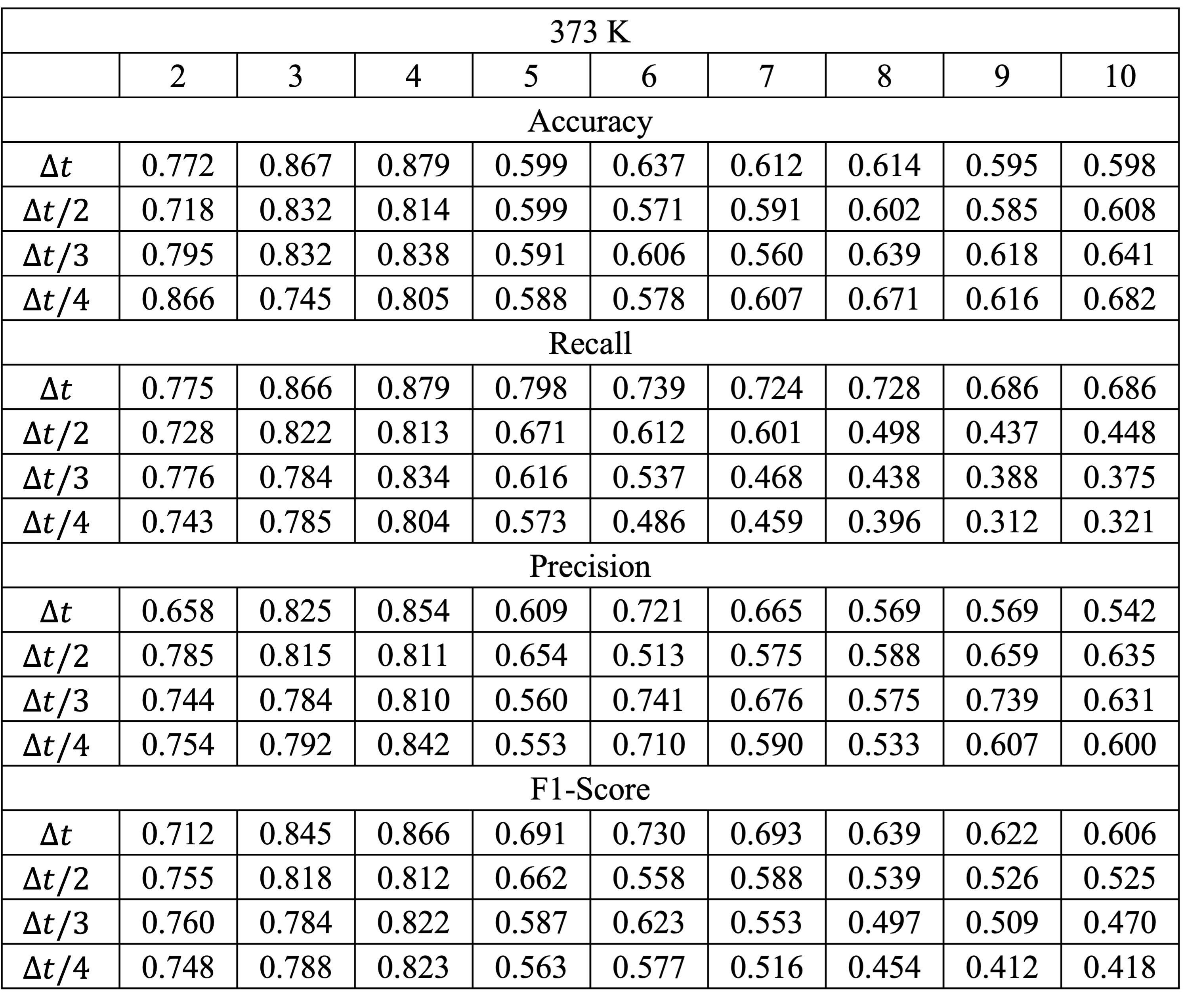}
\end{figure}
\begin{figure}[htbp]
    \centering
    \includegraphics[width=0.5\textwidth]{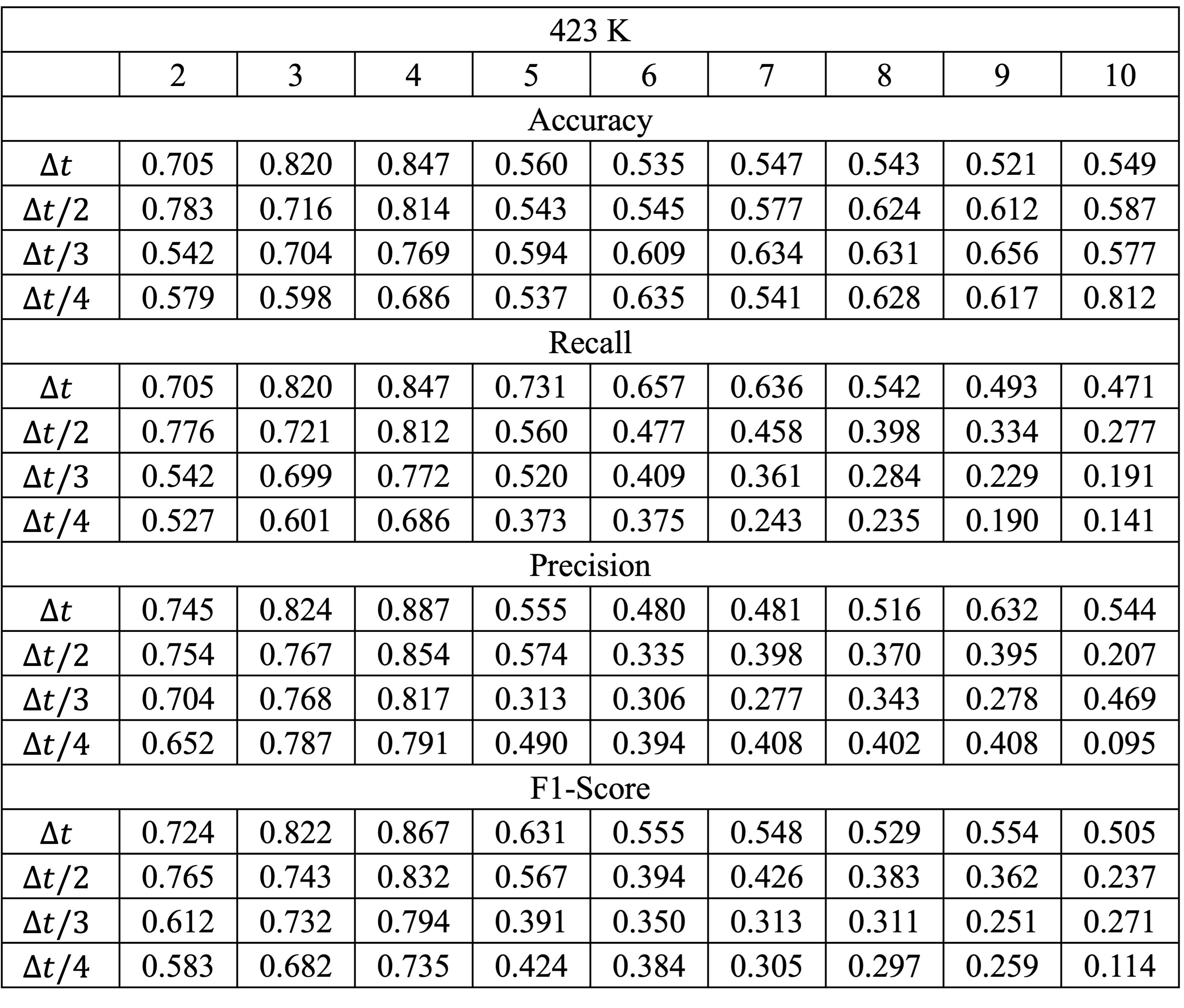}
\end{figure}
\clearpage
\subsection{Lifetime Autocorrelation Functions in Log Scale}
\begin{figure}[htbp]
    \centering
    \includegraphics[width=0.5\textwidth]{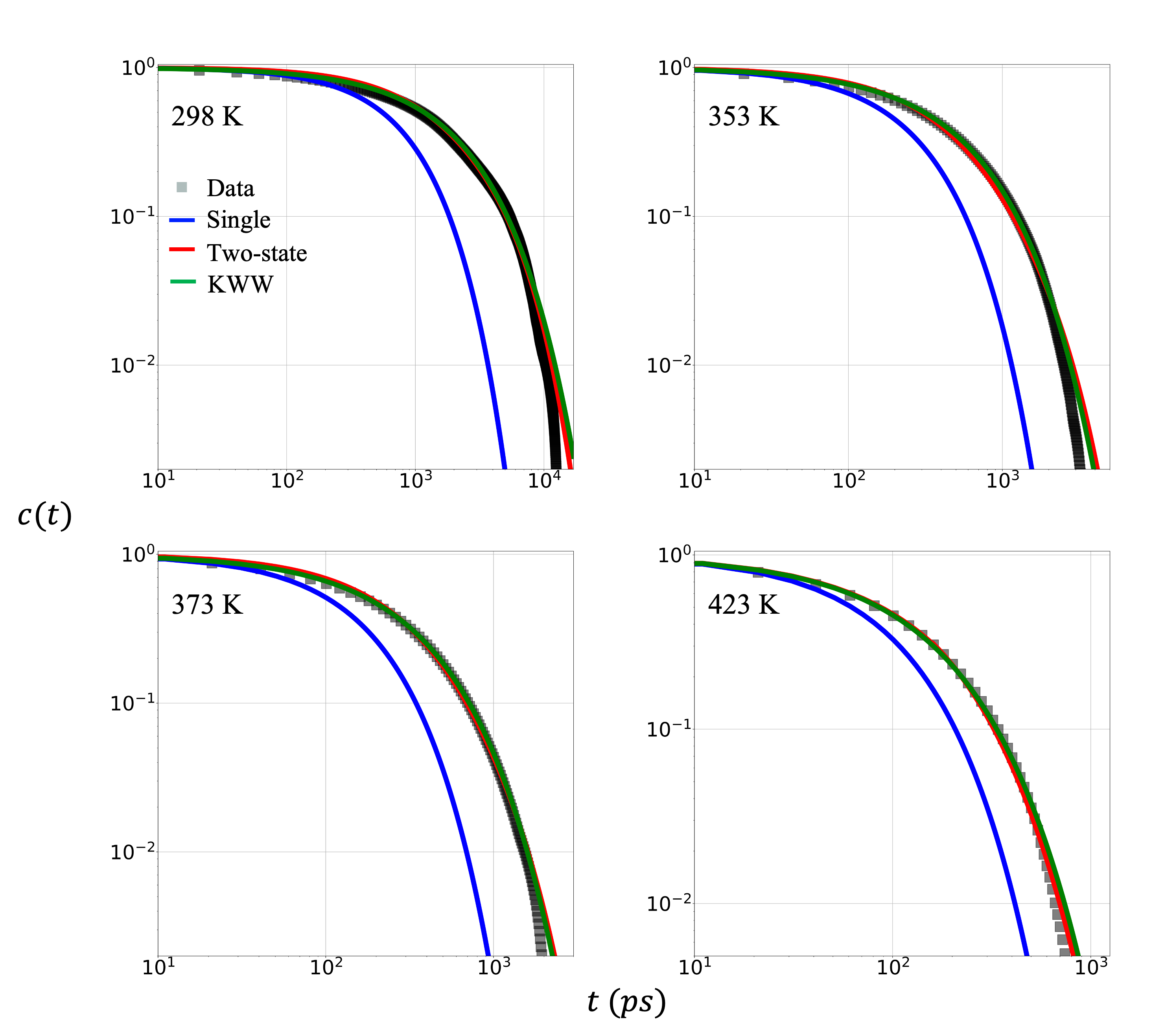}
    \caption{Lifetime autocorrelation functions in log scale; single exponential fitted by $e^{-\mu t}$, Kohlrausch–Williams–Watts (KWW) model (green), and two-state model (red)}
    \label{fig:log_log}
\end{figure}

We compared ACFs with a single exponential, KWW, and our two-state model. A single exponential function only fits a short time scale, and the other two functions fit ACFS well. 

Although the KWW function can also fit the data well, we believe that the two-state model is a more comprehensive approach. Our model considers the microscopic structural changes, which is consistent with the GNN results, and effectively integrates them with the transport properties. Our proposed model offers distinct advantages in understanding the transport mechanism of lithium ions by incorporating the dynamics of microscopic structural changes. 
\newpage
\subsection{Average transition rate}
\begin{table}[ht]
    \centering
    \caption{Average transition rate in each state among temperatures}
    \begin{tabular}{ccc}
        \toprule
        Temperature (K) & $f_{s} (1/ps)$ & $f_{h} (1/ps)$ \\
        \midrule
        298 & 0.149 & 0.00160 \\
        353 & 0.155 & 0.00237 \\
        373 & 0.196 & 0.00442 \\
        423 & 0.239 & 0.00786 \\
        \bottomrule
    \end{tabular}
    \label{tab:data}
\end{table}

We calculated the average transition rate in each state for all temperatures. The average transition rate in the hard state at 423 K is 4.9 times higher than 298 K, while that of the soft state is only 1.6 times higher. Thus, the shell exchange in the soft state is less sensitive than in the hard state.
\end{appendices}


\end{document}